\documentclass[osajnl,twocolumn,showpacs,superscriptaddress,10pt]{revtex4-1}%% use 11pt for Applied Optics
\usepackage{amsmath,amssymb,graphicx,mathtools}
\usepackage{IEEEtrantools}
\usepackage{epstopdf}
\DeclareMathOperator{\sech}{sech}

\begin{document}

\title{Whispering gallery modes in optical fibers based on reflectionless potentials}

\author{Sergey V. Suchkov}\email{Corresponding author: sergey.v.suchkov@anu.edu.au}
\affiliation{Nonlinear Physics Centre, Research School of Physics and Engineering, The Australian National University Canberra, ACT 2601, Australia}

\author{Mikhail Sumetsky}
\affiliation{Aston Institute of Photonics Technology, Aston University, Birmingham B4 7ET, UK}
\author{Andrey A. Sukhorukov}
\affiliation{Nonlinear Physics Centre, Research School of Physics and Engineering, The Australian National University Canberra, ACT 2601, Australia}

\begin{abstract}
We consider an optical fiber with nanoscale variation of the effective fiber radius supporting whispering gallery modes slowly propagating along the fiber, and reveal that the radius variation can be designed to support reflectionless propagation of these modes. We show that reflectionless modulations can realize control of transmission amplitude and temporal delay, while enabling close packing due to the absence of cross-talk, in contrast to conventional potentials.
\end{abstract}


\maketitle %% required

For the past few decades, research and development in photonic integrated circuits has been concentrated on building a platform with miniature dimensions, flexibility, and control needed to deliver breakthrough capabilities in optical computing, communications, and fundamental science. Optical microresonators are indispensable part of such platforms. They have demonstrated great promise as fundamental elements for a variety of applications in photonics and can be implemented for such diverse applications as lasers, amplifiers, sensors, and switchers~\cite{Heebner:OM}.

Two basic platforms for fabrication of photonic circuits with micron-scale elements have been developed: the ring resonator platform~\cite{Xia:2007-65:NP,Vlasov:2008-242:NP,Bogaerts:2010-33:QE,Melloni:2010-181:IEEEPh} and the platform based on photonic crystals~\cite{Joannopoulos:PC}. With the advances in lithographic technology and design methods, these two platforms have steadily increased in complexity, achieved lower loss, and broadened functional capability. However, despite the remarkable accomplishments, the existing platforms still face a severe limitation: it is very difficult to fabricate photonic circuits simultaneously possessing microscopic dimensions and ultra-small losses~\cite{Melloni:2010-181:IEEEPh,Doeer:2006-4763181:IEEEPh}.

An alternative Surface Nanoscale Axial Photonic (SNAP) platform, which combines benefits of microscopic dimensions and ultra-low loss has been proposed in Refs.~\cite{Sumetsky:2012-22537:OE, Sumetsky:2011-26470:OE}. This platform based on an optical fiber with introduced nanoscale modulation of an effective fiber radius (SNAP fiber) including variation of a material refractive index~\cite{Watts:2013-3249:JOSA} and/or a fiber radius~\cite{Sumetsky:2012-22537:OE}. This SNAP fiber supports Whispering Gallery Modes (WGMs) circulating near the surface of the fiber and slowly propagating in the axial direction. Such structure can be considered as a chain of coupled microresonators, which can be close packed to each other allowing precise control of light. The SNAP fiber is coupled to a transverse microfiber, which serves as a source of the WGMs and for detection of transmitted signal. It has recently been shown that this platform can be used for realisation of high-Q factor (intrinsic loss down to $0.44$ dB/ns) dispersionless multinanosecond light traps~\cite{Sumetsky:2014-163901:PRL}.

In this work, we suggest and demonstrate through numerical simulations that the effective fiber radius variation can be designed in a special way to realize so-called reflectionless potentials~\cite{Kay:1956-1503:JAP}. Such modulations transmit all waves with no reflection, which can be used to introduce pure phase advancement with no amplitude change of the WGMs. On the other hand, reflectionless modulations only very weakly affect the optical response at nearby locations, enabling dense packing in contrast to the significant cross-talk associated with ordinary potentials.

\begin{figure}[tb]
    {
        \includegraphics[width=\columnwidth]{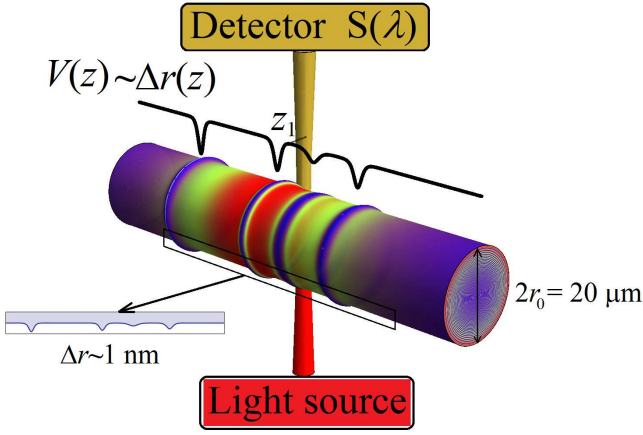}
    }
    \caption{Scheme of a SNAP fiber coupled to a microfiber. The microfiber serves as an input and
    output waveguide and is coupled to the SNAP fiber at position $z_1$. Axial and radial intensity distribution is also presented for $\lambda$=$\lambda_0+0.2 {\rm pm}$ and $z_1=-50 {\rm \mu m}$ for the reflectionless potential $V(z)$ defined by~Eq.(\ref{ReflectionlessPotential}).}
\label{setup}
\end{figure}

We consider a SNAP fiber coupled to a microfiber schematically shown in Fig.~\ref{setup}. The microfiber here is a tapered fiber with a micron-scale diameter waist. The microfiber is coupled to the SNAP fiber at a position $z_1$ and due to an evanescent interaction excites WGMs slowly propagating in the axial direction of the SNAP fiber.

A field distribution of WGMs, adiabatically propagating along $z$-axis of the SNAP fiber can be decomposed into two parts: axial $\Psi$ and radial $\Xi$ as follows: $U(\mathbf{r}) = \Psi(z)\Xi(\rho, \varphi)$, where $(z,\rho,\varphi)$ are the cylindrical coordinates~\cite{Sumetsky:2012-22537:OE}. Here the function $\Psi(z)$ determines light transmission through the microfiber and satisfies a one-dimensional Schr\"{o}dinger like equation with potential $V(z)$, which profile is directly determined by the nanoscale variation of the {\it effective} fiber radius $r_{\rm eff}(z)\sim r(z)n(z)$, where $r(z)$ is a fiber radius and $n(z)$ is medium refractive index. The governing equation for the axial field component is~\cite{Sumetsky:2012-22537:OE}:
\begin{eqnarray}
    \frac{\partial^2 \Psi}{\partial z^2}+\left[E(\lambda)-V(z)+D\delta(z-z_1)\right]\Psi\nonumber\\
    +2ik^2\frac{\gamma}{\lambda_0}\Psi=C\delta(z-z_1),
    \label{modeleq}
\end{eqnarray}
where $E(\lambda)=-2k^2(\lambda-\lambda_0)/\lambda_0$ is an effective energy, $\lambda$ is laser wavelength in vacuum, $\lambda_0$ is a resonant wavelength of the SNAP fiber, $k=2\pi n_0/\lambda_0$ is a propagation constant of light in the optical medium, $n_0$ is a fiber refractive index,  $V(z)=-2k^2\Delta r_{\rm eff}(z)/r_{\rm eff}(z)=-2k^2(\Delta r(z)/r_0+\Delta n(z)/n_0)$ is a potential, $\Delta n(z)$ is refractive index modulation in SNAP, %i.e. $n(z)=n_0+\Delta n(z)$,
$\gamma$ is an attenuation parameter, $z_1$ is the microfiber position, $C$ is a coupling parameter corresponding to energy inflow through the microfiber, and  $D$ determines a phase shift due to the coupling to the microfiber as well as a radiation loss through the microfiber. Since the microfiber diameter waist is much smaller then  characteristic axial wavelength ($1 {\rm \mu m}$ vs. $10 {\rm \mu m}$), the coupling between the microfiber and the SNAP fiber is modeled by means of Dirac delta function. Note that Eq.~(\ref{modeleq}) is valid for an adiabatically slow variation of the fiber radius~\cite{Snyder:1991-OWT}, and we consider this situation in the present work.

The transmission amplitude $S(\lambda)$ (Fig.~\ref{setup}) can be expressed through the Green's function $G(\lambda,z,z_1)$ of the left hand side of Eq.~(\ref{modeleq}) with $D=0$ as follows~\cite{Sumetsky:2012-22537:OE},
\begin{eqnarray}
    S(\lambda)=S^0-\frac{i|C|^2G(\lambda,z_1,z_1)}{1+DG(\lambda,z_1,z_1)},
    \label{TransmissionAmplitude}
\end{eqnarray}
where $S^0$ is a non-resonant component of the transmission amplitude. All parameters, $S$, $D$, and $C$ are slow functions of the wavelength and can be considered as constants in the vicinity of the resonant wavelength $\lambda_0$~\cite{Sumetsky:2012-22537:OE}.

Another important feature of a SNAP device is that it provides an opportunity to arrange a multi-nanoseconds dispersionless time delay of a propagating signal~\cite{Sumetsky:2014-163901:PRL}. The time delay is determined as follows:
\begin{eqnarray}
    \tau(\lambda,z_1)=\frac{\lambda^2}{2\pi c}{\rm Im}\left(\frac{\partial\ln S(\lambda,z_1)}{\partial\lambda}\right).
    \label{TimeDelay}
\end{eqnarray}
Thus, we are able to determine the light transmission amplitude through the microfiber for any arbitrary potential $V(z)$. In the present paper we suggest to use reflectionless potentials which posses nontrivial transmission properties.

An idea of reflectionless potentials  was originally introduced in Ref.~\cite{Kay:1956-1503:JAP}, where an important problem to construct potentials which are fully transparent for electromagnetic waves was addressed. Particularly, the potential of the form
\begin{eqnarray}
    V(z)=V_{\rm ref}(z)\equiv-2\alpha\sech^2[\sqrt{\alpha_1}z],
    \label{ReflectionlessPotential}
\end{eqnarray}
is reflectionless if a depth of the potential well, $\alpha$, is equal to an inverse width of the potential well, $\alpha_1$. Next, we construct a Green's function for this particular potential.

For localized potentials, such that $V(z)\rightarrow0$ at $z \rightarrow \pm \infty$, the Green's function can be constructed as $G(z,z_1)=A(z_1)\Psi_{+}(z)$ for $z \ge z_1$ and $G(z,z_1)=B(z_1)\Psi_{-}(z)$ for $z \le z_1$. The functions $\Psi_{\pm}(z)$ are solutions for the waves propagating away from the microfiber position, accordingly $\Psi_{\pm}(z) = e^{\pm i\sqrt{\tilde{E}(\lambda)}z}$ at $z \rightarrow \pm \infty$. Here $\tilde{E}(\lambda)\equiv E(\lambda)+2ik^2\gamma/\lambda_0$ and square root of $\tilde{E}(\lambda)$ is chosen in the first quarter. To determine the coefficients $A$ and $B$ we use the continuity and the condition for a jump of a derivative of the Green's function at the point $z=z_1$: $A(z_1)\Psi_{+}(z_1)=B(z_1)\Psi_{-}(z_1)$ and  $A(z_1)\Psi_{+}'(z_1)-B(z_1)\Psi_{-}'(z_1)=1$, respectively. Finally, the expression for the Green's function at the point $z=z_1$ is found as
\begin{eqnarray}
    G(z_1,z_1)=\frac{\Psi_+(z_1)\Psi_-(z_1)}{\Psi'_+(z_1)\Psi_-(z_1)-\Psi'_-(z_1)\Psi_+(z_1)}.
    \label{GeneralGreen}
\end{eqnarray}

Next, we analyze a behavior of the Green's function far away from the potential inhomogeneity (well), for concreteness for $z\ge z_0>0$, where $V(z) \simeq 0$. The solutions can be taken as
\begin{eqnarray}
  \Bigg\{
  \begin{array}{l}
  \Psi_+(z)\approx e^{i\sqrt{\tilde{E}(\lambda)}z}, \\
  \Psi_-(z)\approx R e^{i\sqrt{\tilde{E}(\lambda)}z}+e^{-i\sqrt{\tilde{E}(\lambda)}z},
  \end{array}
  \mbox { for } z>z_0.
  \label{ApproxSol}
\end{eqnarray}
Here $R$ is a complex reflection coefficient, which value is defined by scattering from the potential such that $\Psi_-(z)$ satisfies the asymptotic condition at $z\rightarrow -\infty$. The coefficient can be written as $R\equiv|R|e^{i\varphi}$ with real phase~$\varphi$. Substituting~Eq.(\ref{ApproxSol}) into Eq.~(\ref{GeneralGreen}) we obtain
\begin{eqnarray}
  G(\lambda,z_1,z_1)=-i\frac{|R|\exp(i \varphi+2 i \sqrt{\tilde{E}(\lambda)}z_1)+1}{2\sqrt{\tilde{E}(\lambda)}},
  \label{GS}
\end{eqnarray}
and accordingly
\begin{IEEEeqnarray}{l}
  |G(\lambda,z_1,z_1)|^2=\frac{1}{4|\tilde{E}(\lambda)|}\Bigg(|R|^2e^{-4\mathrm{Im}\left[\sqrt{\tilde{E}(\lambda)}\right]z_1}+1\nonumber\\
  +2|R|e^{-2\mathrm{Im}\left[\sqrt{\tilde{E}(\lambda)}\right]z_1}\cos(\varphi+2\mathrm{Re}\left[\sqrt{\tilde{E}(\lambda)}\right]z_1)\Bigg).
  \label{ComplF}
\end{IEEEeqnarray}
It follows from the relation above that if $\mathrm{Re}\sqrt{\tilde{E}(\lambda)}\gg \mathrm{Im}\sqrt{\tilde{E}(\lambda)}$, then there appears a periodic modulation of the transmission amplitude $S$ in $z_1$ direction with the period
\begin{equation}
  T=\pi/\mathrm{Re} \sqrt{\tilde{E}(\lambda)}.
  \label{Per}
\end{equation}
Note that for reflectionless potentials $R=0$, thus the Green's function is non-periodic, as follows from Eq.~(\ref{ComplF}).

For the reflectionless potential we obtain an analytical expression for the Green's function using exact solutions of left hand side of Eq.~(\ref{modeleq}) for $D=0$ with potential~(\ref{ReflectionlessPotential}) derived in~\cite{Kay:1956-1503:JAP} for a conservative case, $\gamma=0$:
\begin{eqnarray}
  \Psi_{\pm}=\frac{\pm i\sqrt{\tilde{E}(\lambda)}-\sqrt{\alpha}\tanh{\sqrt{\alpha}z}}{\sqrt{\alpha}\pm i\sqrt{\tilde{ E}(\lambda)}}e^{\pm i\sqrt{\tilde{E}(\lambda)}z}.
\end{eqnarray}
It is remarkable that these solutions are still valid for the system with losses, $\gamma\neq0$. Thus, the Green's function for reflectionless potential at the point $z=z_1$ is
\begin{eqnarray}
    G(\lambda,z_1,z_1)=\frac{\tilde{E}(\lambda)+\alpha\tanh^2(\sqrt{\alpha}z_1)}{2i\sqrt{\tilde{E}(\lambda)}(\alpha-\tilde{E}(\lambda))}.
    \label{RefGreen}
\end{eqnarray}
This function is non-periodic, in agreement with the discussion above.
\begin{figure}[htb]
    \includegraphics[width=\columnwidth]{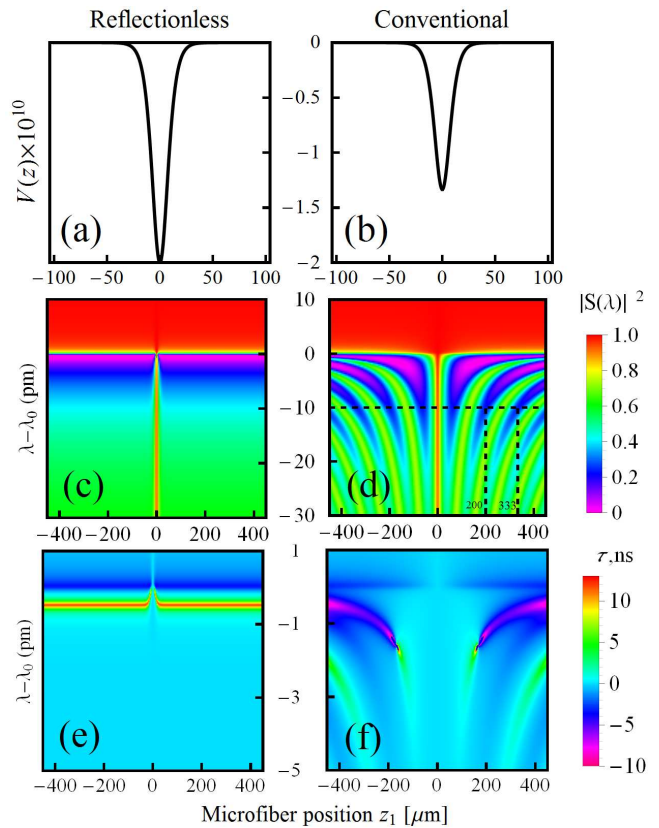}
    \caption{Transmission through the SNAP device based on (a,c,e)~reflectionless potential $V_\mathrm{ ref}(z)$~(\ref{ReflectionlessPotential}) and (b,d,f)~conventional one $V_\mathrm{ con}(z)$~(\ref{ConventionalPotential}). Shown are (a,b)~the potential profiles, (c,d)~transmission amplitude, $|S|^2$, and (e,f)~the time delay, $\tau$, vs. a microfiber position $z_1$ along the SNAP fiber and wavelength deviation around $\lambda_0$.
    In panel (d) dashed lines indicate a modulation period ($\simeq 133\mathrm{\mu m}$) of the transmission amplitude for $\lambda-\lambda_0=-10\mathrm{pm}$.
    }
    \label{Fig1}
\end{figure}

We now compare the transmission characteristics of SNAP devices based on reflectionless potentials with conventional ones of the same width and different depth,
\begin{eqnarray}
    V(z)=V_{\rm con}(z)\equiv-\frac{4}{3}\alpha\sech^2[\sqrt{\alpha_1}z].
    \label{ConventionalPotential}
\end{eqnarray}
For the introduced potentials we determine the transmission amplitude $S(\lambda)$, changing a position of the microfiber along the SNAP fiber, $z_1$, as well as varying the wavelength of the input signal around the resonant value~$\lambda_0$. We use here the following parameter values~\cite{Sumetsky:2012-22537:OE}: $\lambda_0=1.5 {\rm \mu m}$, $\gamma=0.1 {\rm pm}$, $n_0=1.5$, $|C|^2=2\times10^4/\rm{m}$. We choose $\alpha=\alpha_1=0.01 {\rm \mu m^{-2}}$ which corresponds to a $2.53 {\rm nm}$ variation of the fiber radius under assumption of a constant refractive index. We consider a case of lossless coupling between the SNAP fiber and the microfiber, which also does not cause the phase shift of the input light~\cite{Sumetsky:2012-22537:OE}. We checked that the observed effect remains for a lossy coupling as well (when $D\ne i|C|^2/2$). Thus, we put $D=i|C|^2/2$ which leads to $S^0=1$.
Fig.~\ref{Fig1} shows the transmission amplitude $S$ for the SNAP fibers described by reflectionless potential and conventional one. In panels (a-b) the shapes of the potentials are presented, panels (c-d) correspond to the transmission amplitude $|S(\lambda)|^2$, while (e-d) show the time delay $\tau$. The transmission amplitude and time delay are calculated by means of Eqs.~(\ref{TransmissionAmplitude}),(\ref{TimeDelay}). For the reflectionless potentials $V_\mathrm{ ref}(z)$ defined by Eq.~(\ref{ReflectionlessPotential}) we use Eq.~(\ref{RefGreen}), while for the conventional one, $V_\mathrm{con}(z)$, the Green's function is constructed numerically using Eq.~(\ref{GeneralGreen}).

We observe in Fig.~\ref{Fig1} that the reflectionless potential only locally affects the transmission amplitude $|S|^2$, as well as the time delay $\tau$, whilst the conventional potential has an effect far away from the well, resulting in periodical variations of the transmission amplitude. For example, at the wavelength $\lambda=\lambda_0-10\mathrm{pm}$ we have $\mathrm{Re}\sqrt{\tilde{E}(\lambda)}\approx 23000\gg \mathrm{Im}\sqrt{\tilde{E}(\lambda)}\approx115$, and the period according to Eq.~(\ref{Per}) is $T=137\mathrm {\mu m}$, which agrees with numerical results in Fig.~\ref{Fig1}(d).
\begin{figure}[hbt]
    \includegraphics[width=0.9\columnwidth]{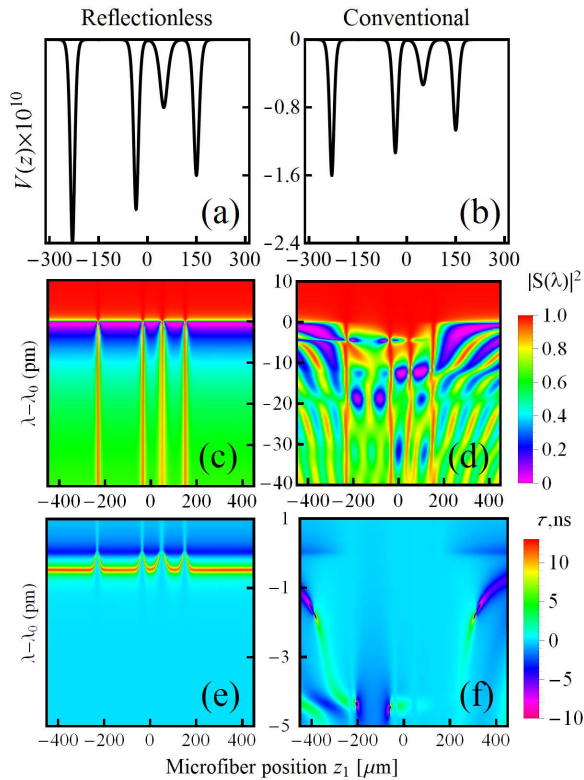}
    \caption{The same as in Fig.~\ref{Fig1}, but for the potentials determined by (a,c,e)~Eq.~(\ref{MultiReflectionless}) and (b,d,f)~(\ref{MultiConventional}).}
    \label{MultPlot}
\end{figure}

We can furthermore construct composite reflectionless potentials satisfying particular requirements. As an example, we consider a superposition of the reflectionless potential wells of the form
\begin{eqnarray}
    V_{\rm mr}(z)\equiv -2 \sum\limits_{j=-2}^{2} \alpha_j \sech^2[\sqrt{\alpha_j}(z+\delta_j)],
    \label{MultiReflectionless}
\end{eqnarray}
with $\delta_j=\{-150,\ -50,\ 35,\ 230\}\ {\rm \mu m}$ and $\alpha_j=\{0.8,\ 0.4,\ 1,\  1.2\}\times10^{-2}\   {\rm \mu m^{-2}}$. This potential is a close packing of the single well reflectionless potentials, so we expect that such potential will also exhibit the reflectionless property. We compare this multi-well potential with conventional one of the form
\begin{eqnarray}
    V_{\rm mc}(z)\equiv\frac23 V_{\rm mr}(z).
    \label{MultiConventional}
\end{eqnarray}
We present the transmission amplitude $S(\lambda)$ and time delay $\tau$ for potentials $V_{\rm mr}$ and $V_{\rm mc}$ in Fig.~\ref{MultPlot}. We use the same parameter values as in the case of single well potentials. From the obtained results it follows that the potential $V_{\rm mr}$ remains reflectionless, while for $V_{\rm mc}$ a strong coupling between the wells is observed, which also causes the modulation of the transmission amplitude vs. the microfiber position.

In summary, we have considered a potential possessing the nontrivial property - that of nonreflection. We have obtained the analytical expression for the transmission amplitude $S(\lambda)$ through the microfiber coupled to the SNAP fiber. We have demonstrated a qualitative difference in light transmission for the reflectionless case compared to the generic one.
Consideration of SNAP resonators utilizing complex types of reflectionless potentials is a promising direction for further investigation.

This work was supported by the Australian Research Council programs, including Future Fellowship FT100100160 and Discovery Project DP130100086. M.~Sumetsky is grateful to the Royal Society and Wolfson Foundation for the Royal Society Wolfson Research Merit Award.

\pagebreak

\end{document}